\documentclass[preprint,12pt]{elsarticle}

\usepackage{amssymb}

\usepackage{url}
\usepackage{amsmath,amsfonts,gensymb}
\usepackage{amssymb,dsfont}
\usepackage{bbm}
\usepackage{caption,subcaption}
\usepackage{booktabs}
\usepackage{float}
\usepackage{graphicx}
\usepackage{xcolor}
\usepackage[normalem]{ulem}

\newcommand{\expec}{\mathbb{E}}
\newcommand{\new}[1]{#1}

\journal{Journal of Theoretical Biology}

\begin{document}

\begin{frontmatter}



\title{Modelling phylogeny in 16S rRNA gene sequencing datasets using string-based kernels}


\author[inst1]{Jonathan Ish-Horowicz}

\affiliation[inst1]{organization={National Heart and Lung Institute},
            addressline={Imperial College London}, 
            city={London},
            postcode={SW7 2AZ}, 
            country={United Kingdom}}

\author[inst2]{Sarah Filippi}

\affiliation[inst2]{organization={Department of Mathematics},
            addressline={Imperial College London}, 
           city={London},
            postcode={SW7 2AZ},
            country={United Kingdom}}

\begin{abstract}
The bacterial microbiome is increasingly being recognised as a key factor in human health, driven in large part by datasets collected using 16S rRNA (ribosomal ribonucleic acid) gene sequencing, which enable cost-effective quantification of the composition of an individual's bacterial community. One of the defining characteristics of 16S rRNA datasets is the evolutionary relationships that exist between taxa (phylogeny). Here, we demonstrate the utility of modelling these phylogenetic relationships in two statistical tasks (the two sample test and host trait prediction) and propose a novel family of kernels for analysing microbiome datasets by leveraging string kernels from the natural language processing literature. We show via simulation studies that a kernel two-sample test using the proposed kernel is sensitive to the phylogenetic scale of the difference between the two populations. In a second set of simulations we also show how Gaussian process modelling with string kernels can infer the distribution of bacterial-host effects across the phylogenetic tree \new{and apply this approach to a real host-trait prediction task.} The results in the paper can be reproduced by running the code at \url{https://github.com/jonathanishhorowicz/modelling_phylogeny_in_16srrna_using_string_kernels}.
\end{abstract}


\begin{highlights}
\item The proposed family of phylogeny-aware kernels leverages string kernels from natural language processing to encode phylogenetic relationships in microbiome datasets.
\item  Simulation studies demonstrate that a kernel two-sample test using one of the proposed string-based kernels is sensitive to the phylogenetic scale at which differences between microbial populations occur, addressing a key limitation of traditional abundance-based only kernels.
\item The proposed kernels can be used within Gaussian Process regression to infer the distribution of bacterial-host phenotype effects across the phylogenetic tree, with validation on both simulated data and a real dataset predicting vaginal pH from bacterial community composition.
\item The relative Gaussian Process training objective using the proposed kernel vs an abundance-only kernel can serve as an indicator of whether the factors controlling a host trait are related to
the observed 16S rRNA gene sequence or if they might be driven by other factors.
\item Open-source code is provided, allowing researchers to replicate all findings and apply the proposed kernels to new microbiome datasets.
\end{highlights}


\end{frontmatter}




\section{Introduction}

\subsection{The human microbiome}

The microbiome is defined as the microorganisms (including bacteria, fungi and viruses), their genetic material and their interactions that live in or on a host organism. The human body is itself a vast and diverse microbial ecosystem, with estimates placing the number of microbial genes per human host at up to ten times larger than the number of human genes \cite{turnbaugh2007human}. Datasets collected via 16S rRNA (ribosomal ribonucleic acid) gene sequencing are driving our rapidly increasing understanding of the role of the microbiome in human health by enabling cost-efficient identification and quantification of bacterial abundance. The 16S rRNA gene region of the bacterial genome has become ubiquitous for bacterial composition analysis as it is universally present in bacterial genomes and contains both conserved and variable regions. Conserved regions make it easy to design primers for polymerase chain reaction primers while variable regions facilitate distinction between different taxa. 

Each variable in a 16S rRNA gene dataset represents a distinct taxon defined by a unique representative sequence. These variables (called operational taxonomic units or OTUs) are related to one another via historical evolutionary relationships (phylogeny) that can be represented by a phylogenetic tree inferred from these representative sequences. These phylogenetic relationships distinguish 16S rRNA gene sequencing datasets from those generated using other sequencing modalities, necessitating tailored statistical methods to appropriately address relevant biomedical questions. In this manuscript, we contribute to the growing literature on non-parametric statistical approaches for analysing such datasets, offering insights into methods that account for their unique structure.


\subsection{Analysis of microbial datasets using kernel methods} 

Kernel methods are popular in biological data analysis as they provide a mechanism by which to encode prior knowledge and can naturally be applied to discrete data types such as sequences (i.e. strings) and trees. \new{In recent years there has been growing interest in kernel-based approaches for a range of microbiome analysis tasks. Kernel regression is probably one of the most common application of kernel methods in the microbial setting, where the primary aim is to test for associations between microbial compositions and clinical labels (e.g. biomarkers or disease status) in a supervised learning framework. These methods include kernel association tests}  \cite{zhao2015testing,wu2016adaptive,koh2017powerful,zhan2018small,koh2019distance,jiang2022mirkat} \new{as well as kernel ridge regression approaches \cite{randolph2018kernel} and, very recently, Gaussian processes \cite{adachi2025quantifying}}. When the clinical label is dichotomous then kernel-based association testing can be performed \new{either using classifiers such as Support Vector Machine \cite{topccuouglu2020framework,ghannam2021machine,nguyen2021associations,li2025best}} or using the kernel two-sample test with the maximum mean discrepancy (MMD, \cite{gretton2012kernel}) as the test statistic \cite{banerjee2019adaptive}.

The choice of kernel function encodes the modelling assumptions of any kernel method. \new{For example, in kernel ridge regression or} Gaussian process regression the kernel determines the characteristics of the regression functions while in the kernel two-sample test it defines the space in which the inner product (similarity) between observations is computed. \new{While various kernels can be used in practice, the radial basis function (RBF) kernel is commonly employed due to its general-purpose flexibility and smoothness properties: given two observations $x, x' \in \mathcal{X}$, where $\mathcal{X}$ is a $p$-dimensional feature space, the radial basis function (RBF) kernel is defined by} $k(x,x') = \sigma^2 \exp \left( -\|x-x'\|^2_2/2l^2 \right)$ where $\sigma^2$ and $l$ are the variance and lengthscale hyperparameters. 
In particular, two phylogenetically similar taxa may have highly conserved functions, ecological niches, pathogenic potential or metabolic pathways. In such cases their host interactions and so their effects on human health may also be similar. 
This motivates the selection of a kernel that incorporates the phylogenetic similarities. \new{Phylogenetic information has previously been incorporated into microbiome data analysis during exploratory or dimensionality reduction stages \cite{fukuyama2012comparisons,fukuyama2017adaptive}, as well as in kernel ridge regression \cite{randolph2018kernel}, and, more recently, in Gaussian Processes \cite{adachi2025quantifying}. This integration is typically achieved by quantifying similarities between biological communities using distance metrics derived from phylogenetic trees \cite{lozupone2005UniFrac,lozupone2007quantitative}.}

\subsection{Our contributions and structure of the paper}
Here we propose a novel family of kernels for microbiome datasets that leverages the fact that each OTU is defined by a representative DNA sequence. The proposed kernel family quantifies the similarity between two samples by defining a distance in terms of the abundance of each OTU in the samples while incorporating information regarding similarities between OTUs. The similarity between the representative sequences of pairs of OTUs is measured using string kernels, which were originally developed in natural language processing for text classification \cite{lodhi2002text} and quickly became popular for the classification of protein sequences in combination with support vector machines \citep{leslie2001spectrum,leslie2003mismatch,leslie2003fast}.

%
We explore the utility of the proposed family of kernels in the context of two important statistical problems: (i) the kernel two-sample test; and (ii) host trait prediction using Gaussian Processes. In particular, through simulation studies, we demonstrate that this family of phylogeny-aware kernels enable a more appropriate kernel two-sample test compared to those that only account for taxa abundance. Furthermore, we illustrate how these kernels can be leveraged within Gaussian Processes to infer the distribution of host phenotype effects across the phylogenetic tree on simulated data as well as on a real dataset to predict vaginal pH from bacterial community composition. The results presented in this paper can be fully reproduced using the code available at \url{https://github.com/jonathanishhorowicz/modelling_phylogeny_in_16srrna_using_string_kernels}. \new{While in this paper we focus on Gaussian Process regression and kernel  two-sample testing, the proposed family of kernels is broadly applicable in the statistical analysis of microbiome datasets and can also be employed in other contexts such as kernel ridge regression and support vector machines, among others.
}

This paper is organised as follows. Section \ref{sec:background} provides an overview of kernel methods, including Gaussian Processes and the kernel two-sample test, before reviewing existing kernels used in microbiome analysis. In Section \ref{sec:string_kernels_for_micro}, we introduce our proposed family of string-based kernels, while Section \ref{sec:somputing_strings} addresses computational considerations for their implementation. The simulation setup for evaluating the performance of these kernels in the context of the kernel two-sample test and host trait prediction using GPs is detailed in Section \ref{sec:simulation}. The results of these simulations are then presented in Sections \ref{sec:two_sample_testing_sim_study} and \ref{sec:gp_sim_study}. Section \ref{sec:gp_real_data} extends the host trait prediction analysis to a real dataset. Finally, we summarize our findings and discuss potential directions for future research in Section \ref{sec:discussion}.

\section{Materials and Methods} \label{sec:materials_and_methods}
\subsection{Background methods}
\label{sec:background}
This manuscript focuses on two statistical tasks that can be performed using kernel-based approaches: (i) a two-sample test using MMD as the test statistic and (ii) supervised learning via Gaussian Processes. The behaviour of kernel methods in both tasks is determined by the choice of a symmetric, positive semi-definite kernel function $k(\cdot,\cdot)$ satisfying
\begin{equation}\label{eq:kernel_trick}
	k(x,x') = \langle \phi(x), \phi(x') \rangle_\mathcal{H}\qquad \forall x,x'\in\mathcal{X} \,,
\end{equation}
for feature map $\phi: \mathcal{X} \rightarrow \mathcal{H}$ which induces a reproducible kernel Hilbert space (RKHS) $\mathcal{H}$. 
In the following section we provide a short introduction on the kernel two sample test and Gaussian Processes before describing previous applications of kernels for microbiome analysis.

\subsubsection{Kernel two sample test: Maximum Mean Discrepancy} 
Given two sets of samples $X=\{x_i\}_{i=1}^{n_x}$ and $Y=\{y_i\}_{i=1}^{n_y}$, where $x_i \overset{\text{i.i.d}}{\sim}P$ and $y_i \overset{\text{i.i.d}}{\sim} Q$, the two-sample test aims to determine which of the two following competing hypotheses best explains the dataset:
\begin{equation}
	H_0 : P = Q \quad\text{v.s.}\quad  H_1: P \neq Q \,,
\end{equation}
with $H_0$ and $H_1$ being called the null and alternative hypotheses respectively. Given a kernel $k(\cdot,\cdot)$, the maximum mean discrepancy (MMD, \cite{gretton2012kernel}) is defined as
\begin{equation} \label{eq:mmd}
	\textnormal{MMD}_k(P,Q) = \| \expec_{x \sim P}[ \phi(x) ] - \expec_{y \sim Q}[ \phi(y) ]\|_\mathcal{H} \;.
\end{equation}
The kernel two-sample test uses as the test statistic the biased, minimum variance estimator \cite{gretton2012kernel} of
\eqref{eq:mmd}, estimated from the samples in $X$ and $Y$:
\begin{equation}
\label{eq:mmd_test_statistic}
\widehat{\textnormal{MMD}}_k^2(X,Y) = \dfrac{1}{n_x^2} \sum_{i,j=1}^{n_x} k(x_i,x_j) +  \dfrac{1}{n_y^2} \sum_{i,j=1}^{n_y} k(y_i,y_j) -  \dfrac{2}{n_x n_y} \sum_{i,j=1}^{n_x,n_y} k(x_i,y_j)  \,.
\end{equation}
 Statistical significance is assessed using a permutation test with $N_\textnormal{perm}$ permutations, and the p-value is given by

\begin{equation}\label{eq:permutation_pvalue}
	p_{\textnormal{perm}} = \dfrac{\sum_{i=1}^{N_\textnormal{perm}} \mathbbm{1}(\widehat{\textnormal{MMD}}_k(X^*_i,Y^*_i) \geq \widehat{\textnormal{MMD}}_k(X,Y)) + 1}{N_\textnormal{perm} + 1} \,,
\end{equation}
where $\{(X^*_i, Y^*_i)\}_{i=1}^{N_\textnormal{perm}}$ is formed by permuting the combined samples of $X$ and $Y$ \cite{phipson2010permutation}.

\subsubsection{Gaussian processes}

Kernel methods can also be used for  non-parametric Bayesian supervised learning tasks via a Gaussian process (GP). Let $X$ be an $n \times p$  input matrix (e.g. containing OTU counts for $p$ OTUs in $n$ samples) and $y=(y_1,\dots y_n)$ an $n$-dimensional host phenotype vector. For a continuous trait, consider the following regression task 
\begin{equation}
y_i=f(x_i)+\varepsilon_i\;, \qquad \varepsilon_i \sim \mathcal{N}(0,\tau^2)\,, \quad i=1\,,\ldots\,,n \,,
\end{equation}
where $x_i$ denotes the $i$-th row of the matrix $X$ and $f(\cdot)$ is an unknown function. To infer this unknown function one can specify a zero-mean GP prior distribution over the function space
\begin{gather} \label{eq:gp_regression_model}
	f(\cdot) \sim \mathcal{GP}(0, k(\cdot,\cdot)) \,,
\end{gather}
which is fully specified by the positive semi-definite kernel function $k(\cdot,\cdot)$ and its hyperparameter(s) $\theta$. The GP prior in \eqref{eq:gp_regression_model} can be seen as a generalisation of a multivariate Gaussian distribution: when evaluating $f(\cdot)$ on a finite set of observations e.g. $x_1,\dots x_n$, the n-dimensional vector $(f(x_1),\dots f(x_n))$ follows a multivariate Gaussian distribution with mean 0 and covariance matrix $K_{XX}$, which is the positive semi-definite matrix with elements formed by pairwise evaluations of $k(\cdot,\cdot)$ on the rows of $X$. The Gaussian likelihood of this regression model permits exact computation of the posterior distribution $p(f(\cdot)\mid X,y)$ via Bayes rule \cite{williams2006gaussian}. In addition, the log-marginal likelihood (LML) of the GP regression model can be obtained analytically
\begin{equation}
    \label{eq:gp-logmarglik}
	\log p \, (y \mid X, \theta) = -\dfrac12 y^T ({K}_{{X}{X}} + \tau^2 {I})^{-1} y - \dfrac12 \log | ({K}_{{X}{X}} + \tau^2 {I}) | - \dfrac{n}{2} \log 2\pi \,.
\end{equation} 
Note that $K_{XX}$ depends on the kernel hyperparameter $\theta$. 

For binary traits, we consider regression models of the form
\begin{equation}
y_i = \Phi(f(x_i))\,, \quad i=1\,,\ldots\,,n \,,
\end{equation}
where $\Phi(\cdot)$ is the cumulative distribution function of the standard Gaussian and $f(\cdot)$ is now a latent function that cannot be inferred in closed-form due to the probit likelihood. In this paper we use the variational GP classifier of \cite{opper2009variational}, which approximates the latent posterior $p(f(\cdot) \mid X, y)$ with a multivariate Gaussian $q(f)=\mathcal{N}(\mu,\Sigma)$ parametrized by $\mu$ and $\Sigma$. The optimal variational distribution $q(\cdot)$ is found by maximising the evidence lower bound (ELBO),
\begin{align} \label{eq:var_gpc_elbo}
    \textnormal{ELBO} =&\, \mathbb{E}_q[ \log p \, (y \mid f, \theta) ] - \textnormal{KL}( \, q(f) \, || \, p(f) \, ) \,, 
\end{align}
with respect to $\mu$, $\Sigma$ and $\theta$, where $\textnormal{KL}( \, q(f) \, || \, p(f) \, )$ is the Kullback-Leibler divergence from $q(f)$ to the prior $p(f)$.

The log-marginal likelihood \eqref{eq:gp-logmarglik} and the ELBO \eqref{eq:var_gpc_elbo} can also be used for model selection (e.g. selection of the kernel and its hyperparameters) respectively in the regression and classification setting \citep{williams2006gaussian,cherief2019consistency}.

\subsubsection{Kernels previously used for microbiome analysis} \label{prev:kernel:microbiome}

The choice of kernel $k(\cdot,\cdot)$ encodes the modelling assumptions of the kernel two-sample test or the GP model and so has a critical effect on their behaviour.  In the two-sample test the choice of kernel function determines the properties of the RKHS $\mathcal{H}$ and so the behaviour of the test statistic in equation~\eqref{eq:mmd}. Meanwhile for a GP, the kernel defines the covariance structure of the prior and so has a strong regularising effect on the functions that can be learnt. 

 \new{Probably the most commonly used kernel in the literature for GP models or when using the MMD test statistic }is the  radial basis function (RBF) kernel defined as
$$ k_\text{RBF}(x,x') = \sigma^2 \exp \left( -\frac{(x-x')^T(x-x')}{2l^2} \right)\;,$$
where $\sigma^2$ and $l$ are respectively the variance and lengthscale hyperparameters, along with other kernels in the Matern family. 
These kernels are said to be \textit{characteristic}, which is known to be a useful property for the kernel two sample test as it guarantees that $\textnormal{MMD}_k(P,Q)=0$ if and only if $P=Q$ \cite{gretton2012kernel}. However, we will illustrate in the next section that in the context of microbiome analysis, where $X$ contains the abundance of each OTU in each sample and does not contain any information related to phylogenetic similarity between OTUs, performing a two sample test or a GP regression using these kernels is not optimal. Indeed, these kernels would ignore any phylogenetic relationships between OTUs.
\new{Despite this, the RBF kernel is still applied to analyse microbiome datasets, especially in the context of SVM classification, kernel regression and the MMD two sample test \cite{banerjee2019adaptive,li2025best,topccuouglu2020framework,nguyen2021associations,ghannam2021machine}.}

\new{In order to incorporate distances or similarities between OTUs in microbiome statistical analysis, previous work has utilized the UniFrac distance ~\cite{lozupone2005UniFrac,lozupone2007quantitative,randolph2018kernel,adachi2025quantifying,fukuyama2012comparisons,fukuyama2017adaptive} - a metric designed to compare biological communities using information from phylogenetic trees.}
Let denote by $x=(x^{(1)}, \dots x^{(p)})\in \mathbb{Z}_{\geq0}^p$ and $x'=(x^{'(1)}, \dots x^{'(p)})\in \mathbb{Z}_{\geq0}^p$ the vectors containing counts for each 
\new{of the $p$ OTUs in the two different samples.  The (unweighted) UniFrac distance between the two samples $x$ and $x'$ is given by the ratio of unshared branch lengths between the two samples to the total branch lengths in the tree:
\begin{equation} \label{eq:unweighted-UniFrac-dist}
	d^\text{uf-uw}(x, x') = \dfrac{\sum_{m} b_m | \mathbbm{1}(A_m(x)>0) - \mathbbm{1}(A_m(x')>0) |}{\sum_{m} b_m \max(\mathbbm{1}(A_m(x)>0), \mathbbm{1}(A_m(x')>0))} \,,
\end{equation}
where $b_m$ is the length of branch $m$, and $A_m(x)$ denotes the numbers of sequences that descend from branch $m$ in the sample $x$ \cite{lozupone2005UniFrac}. The sums on the numerator and on the denominator are taken over all the branches in the phylogenetic tree; and $\mathbbm{1}(A_m(x)>0)$ indicates whether any sequences descends from branch $m$ in sample $x$ or not. A weighted variant of the UniFrac distance allows to weight the branch lengths by the abundances in the two samples and is defined as follows:
\begin{equation} 
	d^\text{uf-w}(x, x') = \sum_{m} b_m \left|\frac{A_m(x)}{A_T(x)} - \frac{A_m(x')}{A_T(x')}\right|\,,
\end{equation}
where $A_T(x)=\sum_{l=1}^p x^{(l)}$ denotes the total number of sequences in sample $x$ 
\cite{lozupone2007quantitative}.}

\new{Given a set of $n$ samples $\{x_1,\dots x_n\} $, we can define the $n \times n$ kernel matrix $K$ (with entries  $K_{ij}=k(x_i,x_j)$) associated to the unweighted UniFrac distance as follows: } $K=-\frac12 J D^\text{uf-uw} J$, where  $D^\text{uf-uw}$ is the $n\times n$ matrix with entries $D^\text{uf-uw}_{ij}=d^\text{uf-uw}(x_i, x_j)$ and  $J = I - \frac{1}{n} 1_n 1_n^T$ is the centring matrix \cite{mardia2024multivariate}. \new{The kernel matrix associated to the weighted UniFrac distance can be defined similarly.}

\new{Note that computing the weighted or unweighted UniFrac distances requires to first infer the phylogenetic tree encoding the evolutionary relationship between the $p$ OTUs as it relies on the knowledge of the branches of the tree and their lengths. }In the next section, we will introduce a new family of string kernels for the analysis of microbiome dataset that directly use the representative sequences of the OTUs instead of inferring the phylogenetic trees to encode similarities between related OTUs.

\subsection{Proposed family of string-based kernels for microbiome analysis} \label{sec:string_kernels_for_micro} 
Here, we propose a novel family of kernels for microbiome datasets that
leverages the fact that each OTU is defined by a representative DNA sequence. The proposed kernel family encodes the similarity between the representative sequences of pairs of OTUs using string kernels commonly used in the natural language or for the classification of protein sequences. In these sequence classification tasks the samples themselves are strings, while in 16S rRNA gene sequencing datasets samples are count vectors whose dimensions (the OTUs) are related to one another by strings (the representative sequences). The proposed family of kernels uses a string kernel to construct an inner product space in which to compute sample-wise similarity. 


Recall that a 16S rRNA gene sequencing dataset consists of a set of vectors $\{x_1,\dots x_n\}$ where each sample $x_i=(x_i^{(1)}, \dots x_i^{(p)})\in \mathbb{Z}_{\geq0}^p$ contains counts for each OTU $l=1, \dots p$. In addition, each OTU is defined by a representative DNA sequence of $\sim$200 base pairs. Denote by $z_l$ the representative DNA sequence for the $l$-th OTU. Let $q(\cdot,\cdot)$ be a string kernel that operates on pairs of strings
\new{and defines a similarity between pairs of representative OTU sequences. This induces a $p\times p$ symmetric matrix of OTU similarities, $(S_q)_{kl} = q(z_k,z_l)$. This matrix $S_q$ is then used to define a quadratic form on the $p$-dimensional microbial abundance vectors as follows: for all $x,x'\in \mathbb{Z}_{\geq0}^p$
$$k_q(x,x')=\langle x, x' \rangle_{S_q} = x^T S_q x'\;.$$
This way, for each kernel $q(\cdot,\cdot)$ operating on pairs of strings, we can define the associated kernel $k_q(x,x')$, which operates on pairs of samples while incorporating OTU similarities via the $S_q$ matrix. If we choose $S_q = I$, where $I$ is the $p \times p$ identity matrix, we are assuming that all OTUs are distinct, resulting in the linear kernel.}
Note that if the vectors of abundances $\{x_1,\dots x_n\}$ are stored in the rows of a $n\times p$ count matrix $X\in\mathbb{Z}_{\geq0}^{n\times p}$, then the \new{sample-wise} kernel matrix $K_q$ associated to the 
kernel $k_q(\cdot,\cdot)$ is a $n\times n$ matrix given by $K_q=X S_q X^T$.

\new{Further intuition on how our proposed kernel encodes phylogenetic similarity between kernels can be obtained by inspecting the $ij^{\textnormal{th}}$ element of $K_q$, which is given by $(K_q)_{ij} = \sum_{k=1}^p \sum_{l=1}^p X_{ik} S_{q,kl} X_{jl}$. The similarity between sample $i$ and sample $j$ is therefore the summed similarity of their abundance of each pair of taxa, weighted by the OTU similarity between the pair of taxa.} 

\begin{figure}
	\centering
	\begin{subfigure}[b]{0.45\columnwidth}
		\centering
		\includegraphics[width=\textwidth]{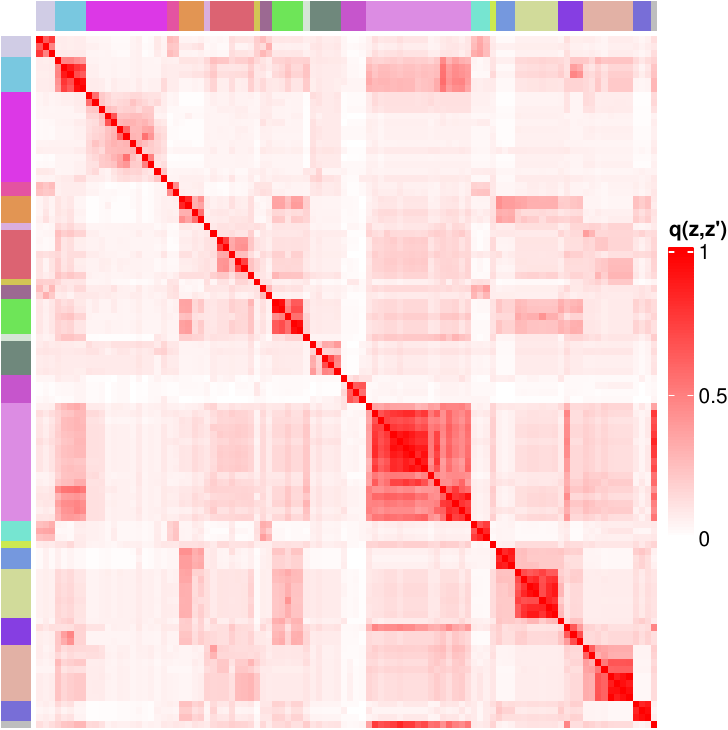}
		\caption{$k=10$}
	\end{subfigure}
	\hspace{2pt}
	\begin{subfigure}[b]{0.45\columnwidth}
		\centering
		\includegraphics[width=\textwidth]{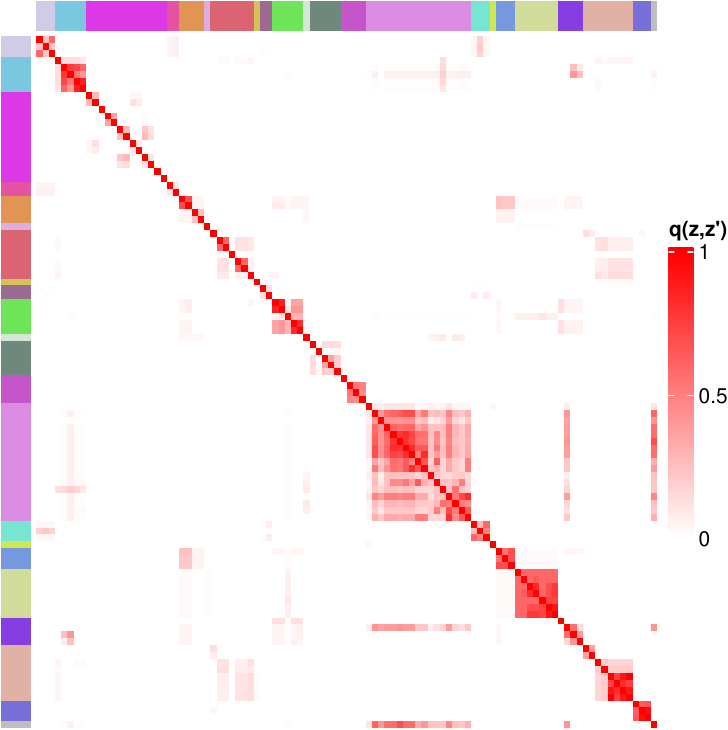}
		\caption{$k=30$}
	\end{subfigure}
	\caption{Spectrum kernels for $k$-mer lengths of 10 (A) and 30 (B). Coloured bars indicate the Order of the OTU, illustrating how blocks of OTUs with high string similarity correspond to taxonomic classifications. The 100 most abundant OTUs from the chronic respiratory disease dataset used in the simulation studies are plotted \cite{cuthbertson2020lung}}.
\label{fig:spectrumkernelvis}
\end{figure}

In the following we consider three string kernels, $q(\cdot,\cdot)$, to quantify similarity between the representative sequences of OTUs: the Spectrum kernel, the Mismatch kernel and the Gappy Pair kernel. The simplest of the three is the Spectrum kernel \cite{leslie2001spectrum}, which is defined by a feature mapping that counts the number of $k$-mers appearing in a string:
\begin{equation}
	\phi^{\textnormal{spec}}(s) = (h^{\textnormal{spec}}_u(s))_{u \in \mathcal{A}^k} \,,
\end{equation}
where $\mathcal{A}^k$ denotes the set of possible $k$-mers in alphabet  $\mathcal{A}$ and $h^{\textnormal{spec}}_u(s)$ returns the number of occurrences of substring $u$ in string $s$. When analysing DNA sequences, $\mathcal{A} = \{\textnormal{T}, \textnormal{G}, \textnormal{C}, \textnormal{A}\}$ corresponding to the four nucleotide and so the $k$-mer feature space $\mathcal{A}^k$ has size $4^k$. 
The Spectrum kernel is then defined for any pairs of representative sequences of OTUs, $z$ and $z'$, 
\begin{equation} \label{eq:string_kernel_k}
	q^{\textnormal{spec}}(z,z') = \langle \phi^{\textnormal{spec}}(z), \phi^{\textnormal{spec}}(z') \rangle_{\mathcal{A}^k},
\end{equation}
 Figure \ref{fig:spectrumkernelvis} illustrates the $S_{q^{\textnormal{spec}}}$ matrices for Spectrum kernels with hyperparameters $k \in \{10,30\}$, computed using the 1,189 OTUs in the respiratory disease dataset utilised throughout this study (described in Section \ref{sec:simulation_studies_overview}, \cite{cuthbertson2022machine}). We observe that smaller values of $k$ produce a matrix with many non-zero elements while larger values of $k$ induce a block diagonal structure, with blocks corresponding to clades of closely-related OTUs.

During replication, DNA sequences undergo mutation, mainly in the form of insertions/deletions (indels) and substitutions. Similarities between two sequences of DNA related through a mutation process would not be recognised by the Spectrum kernel. The Mismatch kernel addresses this by allowing for at most $m$ mismatches when comparing $k$-mers \cite{leslie2003mismatch}. Note that $m$ is an additional hyperparameter whose maximum value is $k-1$. The feature map of the Mismatch kernel is given by 
\begin{equation}
	\phi_m^{\textnormal{mis}}(s) = (h^{\textnormal{mis}}_{u,m}(s))_{u \in \mathcal{A}^k} \,,
\end{equation}
where $h^{\textnormal{mis}}_{u,m}(s)$ counts the number of $k$-mers in string $s$ that have at most $m$ mismatches with $u$.
Another alternative string kernel, called the Gappy Pair kernel, allows for matches between a pair of $k$-mers with up to $g$ gaps, where $g$ is another hyperparameter \cite{leslie2003fast}. Its feature map is
\begin{equation}
	\phi_g^{\textnormal{gap}}(s) = (h^{\textnormal{gap}}_{u,g}(s))_{u \in \mathcal{A}^k} \,,
\end{equation}
where $h^{\textnormal{gap}}_{u,g}(s)$ counts the number of $k$-mers $v$ in string $s$ that matches $u$ with at most $g$ gaps.







To summarise, our proposed family of kernels $k_q(x,x')=x^TS_qx'$ measures the similarity between two samples taking into account the relative abundance of each OTU in the samples (gathered in the vectors $x$ and $x'$) while encoding the similarity between the representative DNA sequences of OTUs through the matrix $S_q$ using the string kernel $q(\cdot, \cdot)$. \new{To better understand the differences and the connections between the proposed kernel and those associated with UniFrac distances, observe that when the count matrix $X$ is column centered, the proposed kernel matrix can be expressed as $K_q=-\frac{1}{2} JD^q J$ where $D^q$ is the $n\times n$ distance matrix with entries $$D^q_{ij}=(x_i-x_j)^TS_q(x_i-x_q)=\sum_{k=1}^p\sum_{l=1}^p (X_{ik}-X_{jk}) S_{q,kl} (X_{il}-X_{jl})\;,$$ where $S_{q,kl}=q(z_k,z_l)$ is the $kl^\textnormal{th}$ element of $S_q$, which is the similarity between OTUs $k$ and $l$. Unlike the weighted and unweighted UniFrac distances, which rely on presence/absence or relative abundances along a phylogenetic tree, the proposed distance incorporates a quadratic form through the matrix, $S_q$, which allows for continuous weighting of pairwise OTU differences. \\ } 

\subsection{Computing String kernels}\label{sec:somputing_strings}

Efficient implementations of string kernels rely on tries, a tree data structure whose leaves represent a set of sequences and where all the children of an internal node have the same prefix \cite{shawe2004kernel}. Tries allow for far more efficient $k$-mer lookups than a naive search in the size of the $k$-mer space, which is exponential in $k$ ($|\mathcal{A}_k|=4^k$). When using tries the time complexity to compute one element in a Spectrum kernel is $\mathcal{O}(k(|z|+|z'|))$ for 
sequences $z,z'$ with lengths $|z|, |z'|$, which is linear in $k$ \citep{shawe2004kernel}.  
The time complexity of the Mismatch kernel is $\mathcal{O}(k^{m+1} |\mathcal{A}_k| (|z|+|z'|))$, which is an increase of $k^m 4^k$ relative to the Spectrum kernel. For a single element of the Gappy pair kernel the running time is $\mathcal{O}(k^g(|z|+|z'|))$, which is an increase by a factor of $k^{g-1}$ relative to the Spectrum kernel \cite{leslie2003fast}.

\begin{figure}
	\centering
	\includegraphics[width=0.7\linewidth]{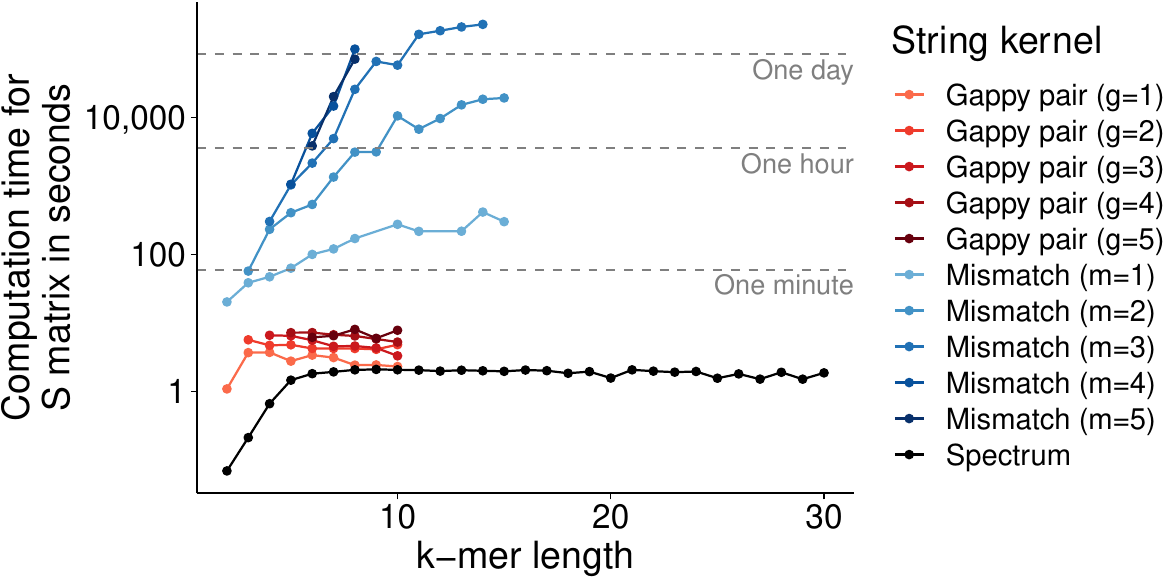}
	\caption{Empirical computation times for the string similarity matrix $S$ for 1,189 OTUs with different hyperparameter values.  Calculations were run on 8 threads of an Intel(R) Xeon(R) CPU using the Kebabs package for R \cite{palme2015kebabs}.}
	\label{fig:kerneltimings_fame}
\end{figure}

The empirical compute times for the same respiratory disease dataset used to produce Figure  \ref{fig:spectrumkernelvis} are shown in Figure~\ref{fig:kerneltimings_fame}, which illustrates that the Mismatch kernel requires at least 3 orders of magnitude more computational time than a Spectrum or Gappy pair kernel for the same $k$-mer length. For the Spectrum, Gappy pair kernels and Mismatch kernels with $m\leq2$ the compute time plateaus once it reaches some value of $k$ (the specific value depends on the type of kernel). This is because for any moderately large $k$ the number of leaves in the trie (which is $4^k$) is far larger than the number of $k$-mers actually present in the two strings $z$ and $z'$, meaning that large parts of the tree are unpopulated. These unpopulated subtrees are pruned before conducting the $k$-mer search and so increasing the value of $k$ does not increase the size of the search in practice \cite{shawe2004kernel}.

While the time complexity of computing string kernels can be restrictive this is mitigated by a combination of two factors. Firstly, the elements of a kernel are independent and so the computational time can be easily reduced using distributed computing infrastructure (so-called embarrassingly parallel computations). Secondly, the nature of microbiome dataset analysis means that the definitions of the OTUs (via their representative sequences) are fixed once the initial pre-processing has been completed. The entire kernel matrix can therefore be computed in advance and stored for future use, and so a computation time on the order of days is feasible as it only has to be performed once. 

\subsection{Simulation study setup}\label{sec:simulation}
In order to demonstrate the impact of the kernel choice and the importance of using kernels that encode similarities between OTUs when applying kernel two sample test or Gaussian Process regression to microbiome dataset, we will devise an appropriate simulation setup as well a real dataset. In this section we describe the simulation set ups.

\subsubsection{Simulating OTU counts} \label{sec:simulation_studies_overview}
Following previous studies, we use a Dirichlet multinomial distribution to generate realistic fictitious OTU count samples \cite{kurtz2015sparse,xiao2018predictive,rong2021mb,patuzzi2019metasparsim,ma2021statistical,gao2017dirichlet}. The Dirichlet multinomial distribution (DMN) is a compound distribution over non-negative integers $\mathbb{Z}_{\geq0}$ that is parameterised by an integer $N \in \mathbb{Z}$ and a vector of concentrations $\alpha \in \mathbb{R}_+^p$  \cite{mosimann1962compound}.  A sample $x \in \mathbb{Z}_{\geq0}^p$ from $\textnormal{DMN}(N,\alpha)$ can be generated as follows:
\begin{equation}
x \sim\, \textnormal{Multinomial}(N, \theta)\quad\text{with}\quad
	\theta=\{\theta_j\}_{j=1}^p\sim\, \textnormal{Dirichlet}(\alpha)\,.
	\end{equation}
In our setting, the number of categories $p$ corresponds to the number of OTUs, while the number of trials $N$ is the total number of reads per sample. To emulate the common scenario where different samples contain different numbers of reads, the number of trials $N \in \mathbb{Z}$ will be different for each sample and be generated using a negative binomial distribution $\textnormal{NB}(a, \, b)$. We use $a=10^5$ and $b \in \{3,10,30\}$ throughout.

To ensure that the generated OTU counts are realistic, the vector of concentrations $\alpha$ is estimated from a real microbiome dataset. Here we use a dataset from the respiratory microbiome of patients with chronic respiratory disease \cite{cuthbertson2022machine}. This contains $p=1,189$ OTUS measured in 107 individuals with cystic fibrosis (83 samples) and non-cystic fibrosis bronchiectasis (24 samples). The collection and preparation of this dataset have been described previously \cite{cuthbertson2020lung,cuthbertson2022machine}. By utilising this real dataset we also have access to its phylogenetic tree, which is inferred from the representative sequences \cite{price2010fasttree} and is used in the simulation set up for the two sample test described in the next subsection. We denote by $\hat\alpha$ the maximum-likelihood estimates of the DMN parameters for this dataset.

\subsubsection{Two sample testing scenario}\label{sec:2sampletestsimu}
In this simulation study we consider two probability distributions,
\begin{gather}
	P = \textnormal{DMN}(N, \alpha) \,, \quad Q = \textnormal{DMN}(N, \tilde\alpha) \,,
\end{gather}
with  $N \sim \textnormal{NB}(10^5, b)$, meaning that the difference between $P$ and $Q$ is fully defined by the difference between the vectors of concentrations $\alpha$ and $\tilde\alpha$. The aim of this simulation study is to demonstrate that only phylogenetic-aware kernels offer two-sample tests that are sensitive to the phylogenetic scale of the difference between $P$ and $Q$. To do this we will fix $\alpha=\hat\alpha$ and vary $\tilde\alpha$ by controlling its distance to $\alpha$ according to a given phylogenetic tree.

\begin{figure}
	\centering
	\includegraphics[width=\linewidth]{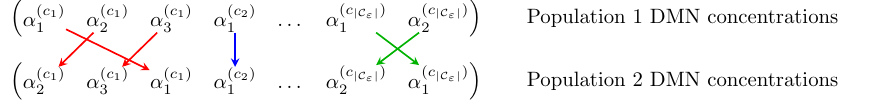}
	\caption{The difference between the two populations in the two-sample test simulation study is a permutation that restricts swaps to those within a set of clusters $\mathcal{C}_\varepsilon = \{c_1 \,,\, \ldots \,,\, c_{|\mathcal{C}_\varepsilon|}\}$. Here $\alpha_i^{(c_k)}$ is the DMN concentration of the $i^\textnormal{th}$ OTU in cluster $c_k$. In this example the clusters $c_1$, $c_2$ and $c_{|\mathcal{C}_\varepsilon|}$ have sizes 3, 1 and 2 respectively.}
	\label{fig:otu_perm_equation_fig}
\end{figure}

In a nutshell, we will define $\tilde{\alpha}$ as being a vector of length $p$ whose elements are a permutation of the elements in $\alpha$. The permutation will be restricted to only swapping elements corresponding to OTUs that are phylogenetically similar, with the level of similarity between OTUs being determined by the phylogenetic tree and a hyperparameter $\varepsilon$. If $\varepsilon=0$, every OTU will be considered independent and $\tilde\alpha$ could be any permutation of $\alpha$. However, a value of $\varepsilon>0$ will define a partition of the OTUs where similar OTUs according to the phylogenetic tree are grouped together. The permutation will therefore only swap elements of $\alpha$ with a maximum phylogenetic similarity proportional to $\varepsilon$. The effect of the permutation on the DMN concentrations is illustrated in Figure \ref{fig:otu_perm_equation_fig} for a toy example.

We now describe this process in more details. Note that for any $\varepsilon>0$, there exists a partition $\mathcal{C}_\varepsilon = \{c_1 \,,\, \ldots \,,\, c_{|\mathcal{C}_\varepsilon|}\}$ of the set of OTUs $\{1,\dots p\}$ that satisfies
\begin{equation} \label{eq:otu_cluster_def}
	\forall c_k \in \mathcal{C}_\varepsilon \quad\forall i,j \in c_k \quad\Delta^\tau_{ij} \leq \varepsilon \Delta^\tau_{\max} 
\end{equation}
where $\Delta^\tau_{ij}$ is the distance between OTUs $i$ and $j$ along the branches of the phylogenetic tree and $\Delta^\tau_{\max}$ is the maximum distance between any two OTUs.  We now define a set of functions $\pi_\varepsilon:\mathbb{Z}^p_{\geq 0}\to\mathbb{Z}^p_{\geq 0}$ such that for all $\alpha\in \mathbb{Z}^p_{\geq 0}$ and for all $c_k \in \mathcal{C}_\varepsilon$
$$\{(\pi_\varepsilon(\alpha))_j,\; j\in c_k\}=\{\alpha_j,\; j\in c_k\}\;. $$
Therefore setting $\tilde\alpha=\pi_\varepsilon(\alpha)$ ensures that for every set of OTU $c_k$ the set of concentrations associated to these OTUs is the same in $P$ and $Q$. The specific OTUs to which a concentration is assigned may differ between $P$ and $Q$ if $c_k$ contains more than one item.  As the partition $\mathcal{C}_\varepsilon$ is constructed based on the phylogenetic distances between OTUs then the difference between $P$ and $Q$ is directly related to the phylogenetic scale. In the simulation study, we use the phylogenetic tree associated to the dataset described in Section~\ref{sec:simulation_studies_overview}. 

The scale of phylogenetic differences between the two populations is controlled by $\varepsilon$. In this simulation study we consider $\varepsilon \in \{0, 10^{-2}, 10^{-1}, 1\}$, where $\varepsilon=0$ corresponds to the null hypothesis where $P=Q$.



\subsubsection{Gaussian Processes regression scenario}\label{sec:simGP}
In the second simulation study we aim to simulate a host trait prediction scenario where the host phenotype is related to the OTU abundances. Again, we simulate OTU abundances $X \in \mathbb{Z}_{\geq0}^{n \times p}$ using a $DMN(N, \hat\alpha)$ where $\hat\alpha$ is the maximum likelihood estimate for the vector of concentrations from the chronic respiratory disease dataset, $N=10^5$ and $n \in \{25, 50, 100, 200\}$. 

In this section we simulate both continuous and binary host phenotypes. We follow \cite{xiao2018predictive} and assume that the relative abundance of each OTU in a sample is the relevant quantity when determining host phenotype. More precisely, given the simulated OTU counts $X$, a fictitious continuous host phenotype $y \in \mathbb{R}^n$ is generated from the relative abundances $Z \in [0,1]^{n \times p}$ where $Z_{ij} = \frac{X_{ij}}{\sum_k{X_{ik}}}$ using a linear model of the form
\begin{equation} \label{eq:phenotype_model}
	y = \beta Z  + \eta \,, \quad \eta \sim \mathcal{N}(0, \rho^2I) \,,
\end{equation}
where $\beta \in \mathbb{R}^p$ are the effect sizes. The variance of $\beta Z$ is fixed to 1 throughout and we consider two noise-levels defined by $\rho \in \{0.3, 0.6\}$, corresponding to signal to noise ratios of $\frac{10}{3}$ and $\frac{10}{6}$. 
Similarly, a fictitious binary host phenotype can be generated using the following thresholded-version of \eqref{eq:phenotype_model}:
\begin{equation} 
	y = \mathbbm{1}(\beta Z  + \eta \geq 0) \,, \quad \eta \sim \mathcal{N}(0, \rho^2I) \,.
\end{equation}
Here we fix $\rho^2=0.1$.

The phylogenetic component of the simulation is introduced via the OTU effect sizes $\beta$, which are assigned to clusters of OTUs in two scenarios, each of which represents a distinct biological hypothesis:

\begin{itemize}
	\item \underline{Scenario 1:} OTU effects are driven by the 16S rRNA gene sequence and so phylogenetically similar OTUs have similar effects; 
	\item \underline{Scenario 2:} OTU effects are assigned at random and are unrelated to the tree and 16S rRNA gene sequence.
\end{itemize}

For both scenarios the set of OTUs $\{1,\dots p\}$ is partitioned and we assign values of the effect size to each OTU according to this partition as follows: given a partition $\mathcal{C}=\{c_1,\dots c_K\}$ of the set of OTUs, we first randomly sample without replacement a subset $\{c'_1, \dots c'_{10}\}$ of  $\mathcal{C}$ and then sample $\tilde{\beta} \sim \mathcal{N}(0, 10 \, I_{10})$; the OTU-level effects are then given by
\begin{align}\label{key}
	\beta_j = \begin{cases}
		\tilde{\beta}_k & \textnormal{if OTU } j \textnormal{ is in cluster } c'_k \\
		0 & \textnormal{otherwise} \\
	\end{cases}  \quad j=1, \ldots, p\;.
\end{align}
This results in a sparse vector $\beta$ with ten unique values.

\begin{figure}
	\centering
	\begin{subfigure}[b]{0.45\columnwidth}
		\centering
		\includegraphics[width=\textwidth]{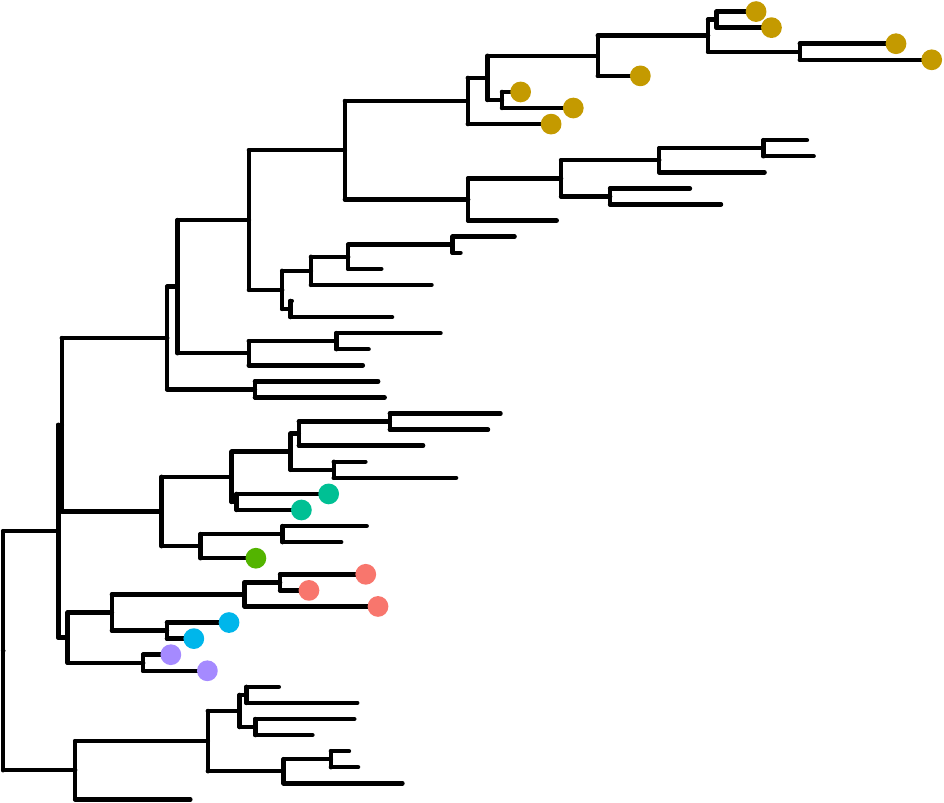} 	
		\caption{Effect sizes follow tree}
	\end{subfigure} 
	\begin{subfigure}[b]{0.45\columnwidth}
		\centering
		\includegraphics[width=\textwidth]{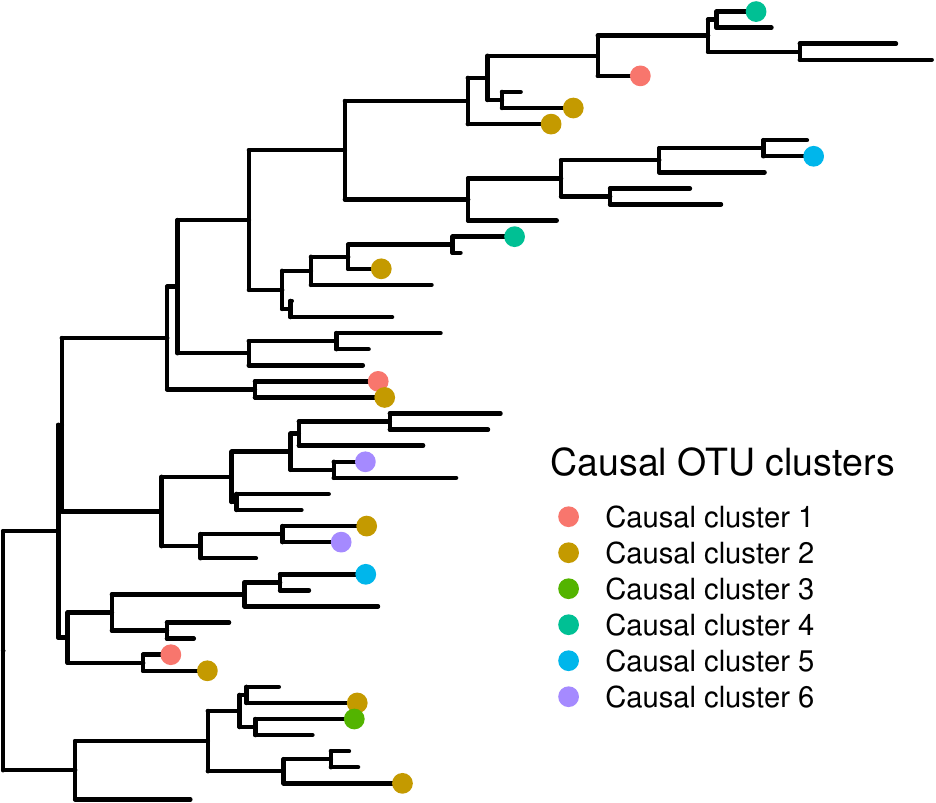}
		\caption{Effect sizes unrelated to tree}
	\end{subfigure}
	\caption{Generating OTU effect sizes that are related to phylogeny (plot A) or are unrelated to phylogeny (plot B). Unmarked leaves denote OTUs with zero effect size in the phenotype model.} 
	\label{fig:supervised_effect_sizes}
\end{figure}

For Scenario 1 the set of OTUs is partitioned according to the phylogenetic tree as described in  Section~\ref{sec:2sampletestsimu} to obtain the partition $\mathcal{C}_\varepsilon=\{c_1,\dots c_{|\mathcal{C}_\varepsilon|}\}$ with $\varepsilon=0.1$. This allows OTUs that are closely related according to the phylogenetic tree to have the same effect size. For Scenario 2, each OTU is randomly allocated to a set $c_k\in\mathcal{C}_\varepsilon$ to form a partition of the OTUs with the same number of clusters and clusters of same size than in Scenario 1 but where OTUs are clustered unrelated to their phylogenetic relationship.  The distribution of OTU effect sizes in the two scenarios is illustrated in Figure \ref{fig:supervised_effect_sizes}. 

\section{Results}

\subsection{Simulation study I: Two-sample testing} \label{sec:two_sample_testing_sim_study}


Consider two samples $X =\{x_i\}_{i=1}^{n_x}$ and $Y =\{y_i\}_{i=1}^{n_y}$ simulated as described in section~\ref{sec:2sampletestsimu} and briefly summarised here:
\begin{align}
\label{eq:twosample_sim_start} 
	X =&\, \{x_i\}_{i=1}^{n_x} \sim \textnormal{DMN}(N, \alpha_1), \\
	Y =&\, \{y_i\}_{i=1}^{n_y} \sim \textnormal{DMN}(N, \alpha_2), \quad \text{with }
	\alpha_2 = \pi_\varepsilon(\alpha_1) \,, \\
    N \sim&\, \textnormal{NB}(10^5, b), \label{eq:twosample_sim_end}
\end{align}
where the scale of phylogenetic differences between the two populations is controlled by $\varepsilon$. We consider $\varepsilon \in \{0, 10^{-2}, 10^{-1}, 1\}$, where $\varepsilon=0$ corresponds to the null hypothesis and $\varepsilon=1$ corresponds to a single cluster containing all $p$ OTUs. Throughout these experiments $n_x=n_y=n \in \{25,50,100,200\}$ and $b=10$.


The aim of the study is to investigate the behaviour of the two-sample test with $\widehat{\textnormal{MMD}}_k(X,Y)$ as the test statistic. An appropriate kernel, $k$, induces a two-sample test which has well-calibrated Type I error and high power, but is also sensitive to the value of $\varepsilon$. Recall that $\varepsilon$ is a simulation parameter that expresses the minimum phylogenetic similarity between differentially expressed taxa. In this study we compare the performance of the test using 
\begin{itemize}
    \item the proposed kernels $k_q(\cdot,\cdot)$ with three different choices of string kernels: (i) the Spectrum kernel with $k \in \{2, \ldots, 30\}$, (ii) the Mismatch kernel with $k \in \{2, \ldots, 15\}$ and $m \in\{1,2,3,4,5\}$, and (iii)  the Gappy pair kernel with $k \in \{2, \ldots, 15\}$ and $g \in\{1,2,3,4,5\}$
    \item the weighted and unweighted UniFrac kernels
    \item and two abundance-only kernels: the RBF kernel with median heuristic lengthscale \cite{garreau2017large}, and the linear kernel defined as $k(x,x')=x^Tx'$. 
\end{itemize}


We generated 100 datasets using \eqref{eq:twosample_sim_start}-\eqref{eq:twosample_sim_end} and used the fraction in which $H_0$ is rejected to evaluate the behaviour of the two-sample test with a given kernel. In each replicate we set $\alpha_1$ to be a permuted version of the Maximum likelihood estimates $\hat\alpha$ of the DMN concentrations for the chronic respiratory disease dataset described in Section \ref{sec:simulation_studies_overview}. We use a nominal significance level of 0.1, for which a well-calibrated test rejects $H_0$ close to 10\% of the time when data are simulated under the null hypothesis. When $\varepsilon=0$ (i.e. $P=Q$), if the observed rate of $H_0$ rejections is significantly different from 10\% then the Type I error of the test is poorly-calibrated.  When $\varepsilon>0$, $P\neq Q$ and so a higher rate of $H_0$ rejections indicates higher power.

\begin{figure}
    \centering
    \begin{subfigure}{\columnwidth}
        \centering
        \includegraphics[width=\columnwidth]{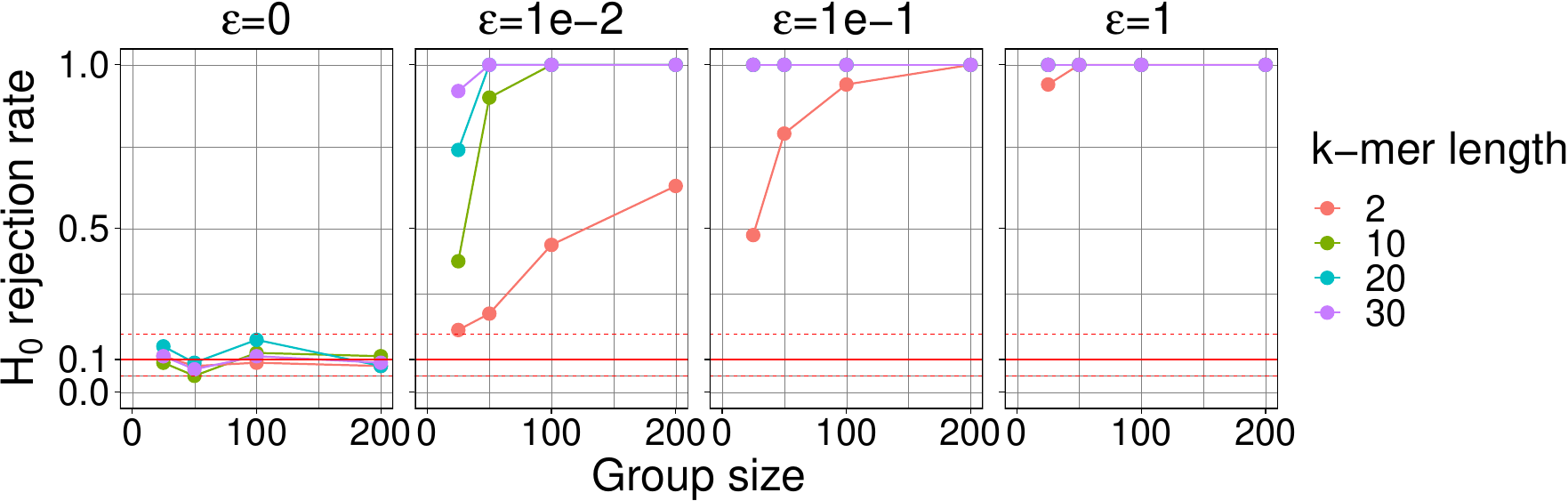}
        \caption{Spectrum kernels}
    \end{subfigure}
    \begin{subfigure}{\columnwidth}
        \centering
        \includegraphics[width=\columnwidth]{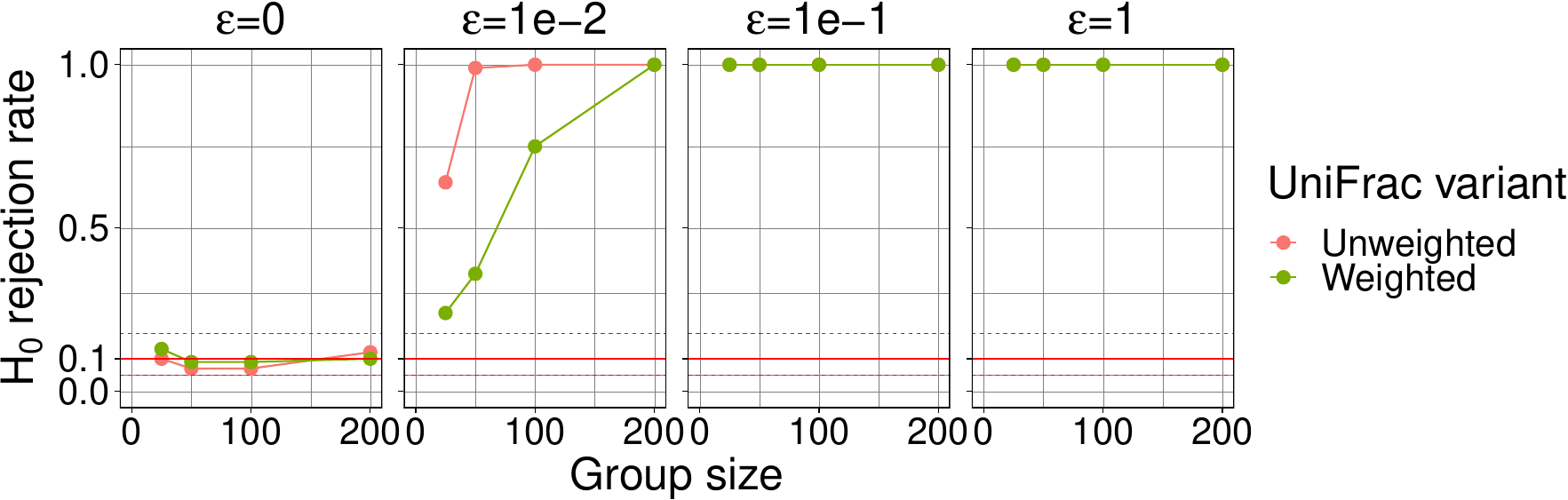}
        \caption{UniFrac kernels}
    \end{subfigure}
    \begin{subfigure}{\columnwidth}
        \centering
        \includegraphics[width=\columnwidth]{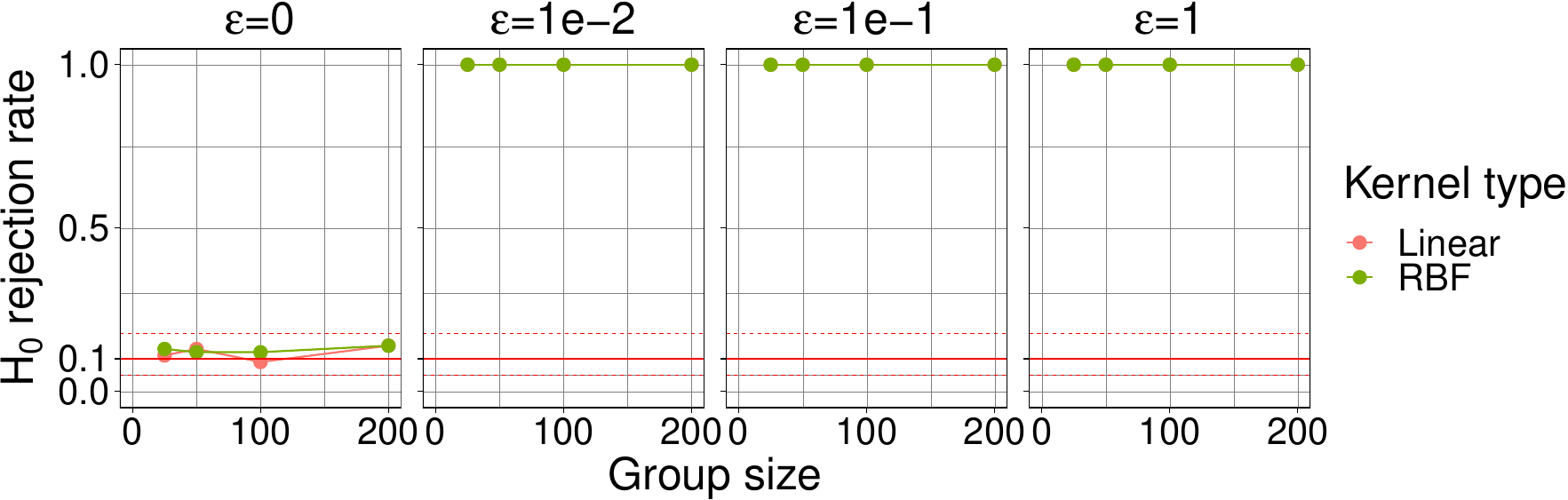}
        \caption{Abundance-only kernels}
    \end{subfigure}
    \caption{Rate of null hypothesis rejections in the two-sample test simulation study for: (A) the proposed kernel using the Spectrum string-kernel, (B) UniFrac kernels and (C) abundance-only kernels. The solid red line denotes the nominal significance level (0.1) and the dashed lines show its 95\% binomial proportion confidence interval.}
    \label{fig:mmd_H0_rej_rate}
\end{figure}

Figure \ref{fig:mmd_H0_rej_rate} shows the $H_0$ rejection rate for the proposed kernel using the Spectrum string-kernel (top row), the Unweighted and Weighted UniFrac kernels (middle row) and the two abundance-only kernels (bottom row). We observe that all kernels induce a test with well-calibrated Type I error (left-hand column, $\varepsilon=0$). \new{When $\varepsilon=10^{-2}$ the proposed Spectrum string kernel with $k \in \{20, 30\}$ has a higher power than both the weighted and unweighted UniFrac kernels. For $k=10$ the power is higher than the weighted UniFrac kernel, but lower than the unweighted UniFrac kernel, while both UniFrac kernels have higher power than our proposed kernel with $k=2$. For $\varepsilon \in \{10^{-1}, 1\}$ the proposed Spectrum kernel with $k>2$ and both UniFrac kernels have power close to 1.}


The abundance-only kernels at first glance may seem to be the optimal choice as they have the highest power. However, this is actually a drawback as they are overly sensitive to differences between $P$ and $Q$ that may not have biological relevance. These kernels do not model any phylogenetic relationships and weight all differences between OTUs equally. They are therefore very likely to reject $H_0$ based on differences between very closely-related (and often indistinguishable) OTUs. As stated previously, an appropriate two-sample test for microbial applications should be sensitive to the phylogenetic scale on which $P$ and $Q$ differ.

\begin{figure}
    \centering
    \begin{subfigure}{\columnwidth}
        \centering
        \includegraphics[width=\columnwidth]{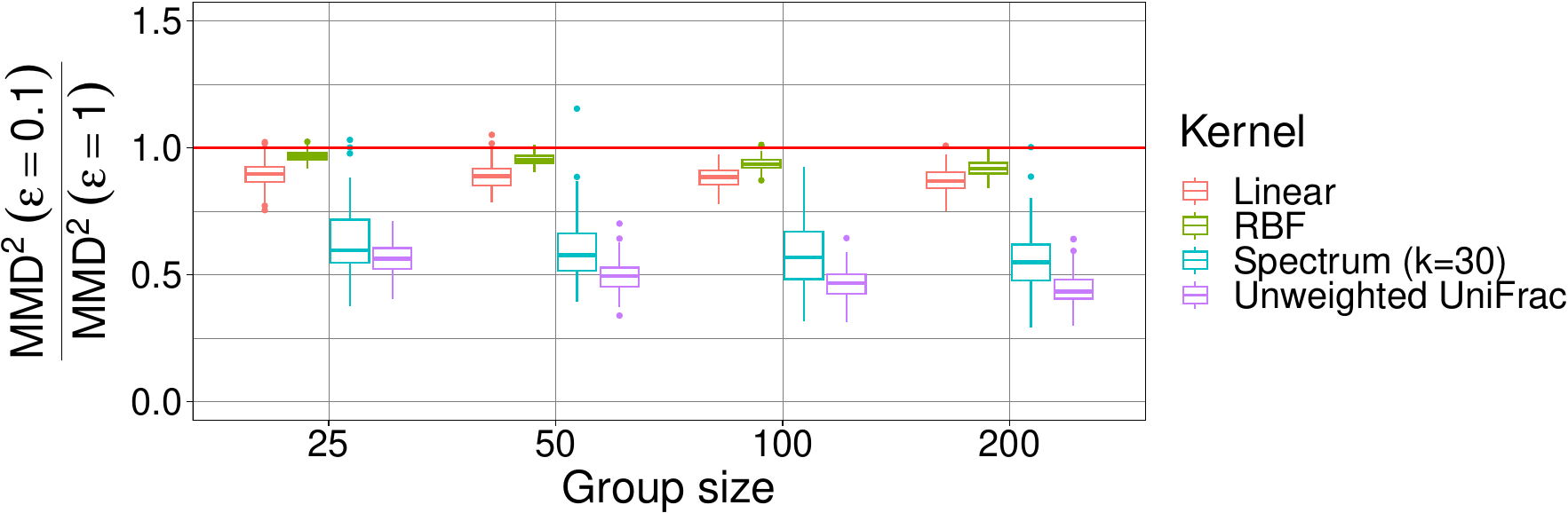}
        \caption{}
    \end{subfigure}
    \begin{subfigure}{\columnwidth}
        \centering
        \includegraphics[width=\columnwidth]{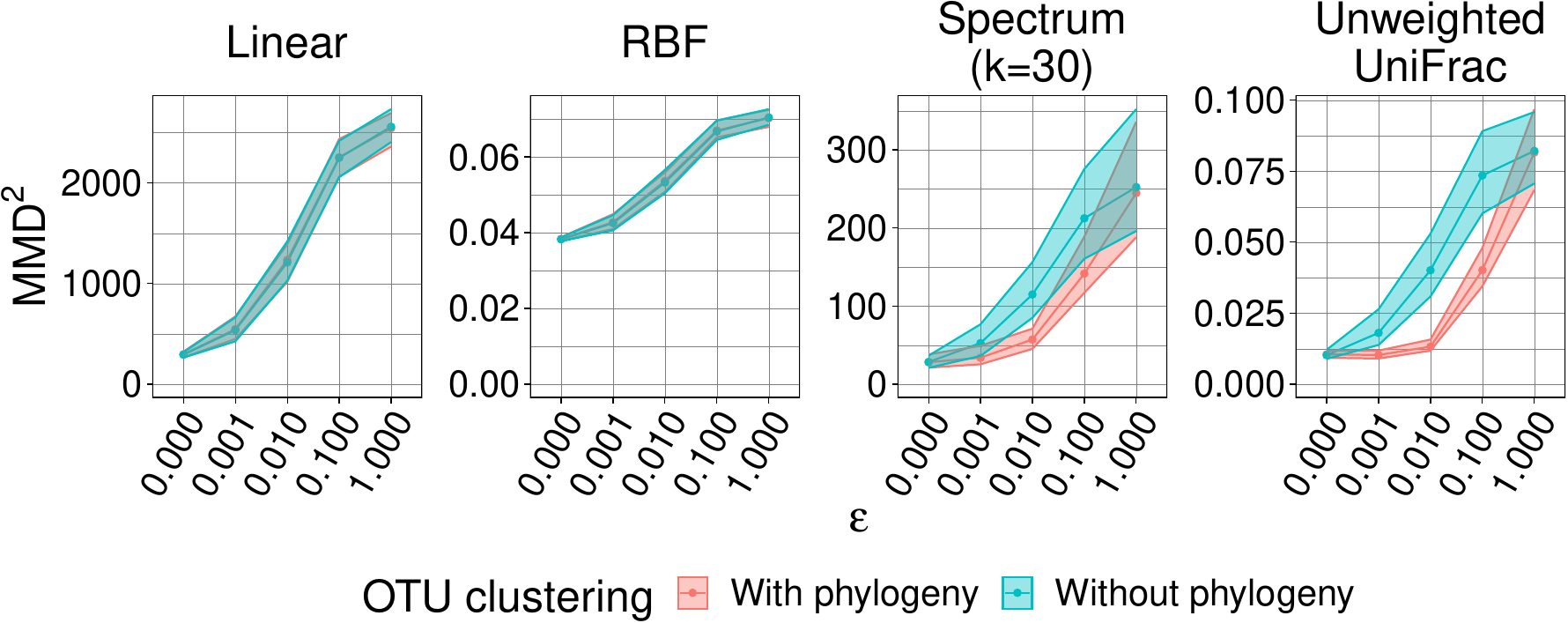}
        \caption{}
    \end{subfigure}
    \caption{(A): the ratio between $\textnormal{MMD}^2(X,Y)$ when $\varepsilon=0.1$ and $\varepsilon=1.0$ shows that the kernels that only model OTU abundances have similar MMD values for very different phylogenetic scenarios, while phylogenetic kernels (spectrum-30 and unweighted UniFrac) have far lower MMD values when $\varepsilon=0.1$. (B): defining OTU clusters without using phylogeny does not change the MMD values for abundance-only kernels.)}
    \label{fig:mmd_phylo_study}
\end{figure}

For a single replicate the DMN concentrations $\alpha_1$ are fixed and $\alpha_2=\pi_\varepsilon(\alpha_1)$ for a function $\pi_\varepsilon(\cdot)$ using a sequence of increasing $\varepsilon$ values. Therefore, an appropriate RKHS for microbiome applications should produce larger MMD values when $\varepsilon=1$ than when $\varepsilon=0.1$. The two scenarios represented by these values of $\varepsilon$ are very different, as $\varepsilon=1$ imposes no phylogenetic restrictions on the differences between the probability distributions $P$ and $Q$, but $\varepsilon=0.1$ forces any differences to occur amongst OTUs that are at most 10\% of the total phylogenetic variation apart. Figure \ref{fig:mmd_phylo_study}(A) shows that the MMD value when $\varepsilon=0.1$ is far smaller than its value when $\varepsilon=1$ for the Spectrum $k=30$ and Unweighted UniFrac kernel. 

Figure \ref{fig:mmd_phylo_study}(A) also suggests that the Linear and RBF kernels produce smaller MMD values when $\varepsilon=0.1$ than when $\varepsilon=1$, although not to the same degree. We now show that this difference in MMD is unrelated to phylogeny. To do so, we compare the MMD when $\alpha_2$ is computed using a set of clusters with the same sizes as $\mathcal{C}_\varepsilon$, but whose labels are assigned at random (without using the phylogenetic tree). The result is a set of permutations with the same properties as $\pi_\varepsilon$ but that have no relation to phylogeny.
Figure \ref{fig:mmd_phylo_study}(B) compares MMD values calculated when $\alpha_1$ and $\alpha_2$ are related to one another by permutations with and without phylogenetic information. MMDs for the proposed kernel using the Spectrum string-kernel  ($k=30$) and Unweighted UniFrac kernels have distinct MMD distributions between the two scenarios, but abundance-only (Linear and RBF) kernels have identical distributions. \new{Note that Figure \ref{fig:mmd_phylo_study} shows the dependence of the MMD test statistic on $\varepsilon$ and does not consider the power of the test.}


\new{As observed along this section, the choice of the value of the hyperparameter $k$ affects the performance of the kernel two-sample test; in particular, larger values of $k$ increase the power. Given this is a single simulation study using counts simulated from a single dataset we cannot give a rigorously tested recommendation on how to choose $k$ a priori. However, based on these results we suggest using the largest possible value of $k$ in practice as a reasonable heuristic. This corresponds to setting the similarity of pairs of taxa to zero if they do not have a recent common ancestor (see Figure \ref{fig:spectrumkernelvis}). If the test is being run as part of an exploratory analysis (with less concern for multiple testing) then we recommend conducting a sensitivity analysis of the p-values with respect to the choice of $k$.}

In conclusion, this simulation study demonstrates that the proposed kernel using the Spectrum string-kernel offers higher power than UniFrac kernels, while still modelling phylogenetic features of microbial datasets.

\subsection{Simulation study II: Host trait prediction using Gaussian processes} \label{sec:gp_sim_study}


Host trait prediction is another important task in microbial studies. The aim of this set of simulations is to identify scenarios under which a phylogenetic-aware kernel improves the training data fit and predictive performance of a Gaussian Process model.

We generate 100 datasets for each of the six simulation setups described in section~\ref{sec:simGP}, i.e. two regression models with different levels of additive noise as well as one classification model; with effect size either generated so that they are affected by the phylogeny or not. For each of the datasets, GP models are trained using a linear and one of the proposed string-based kernels. \new{Note that the proposed family of string kernels reduces to the linear kernel when the matrix $S_q$ is set to the identity; therefore, comparing GP models fitted with these two kernels allows us to investigate whether variation in the host trait is associated with the structure of the observed 16S rRNA gene sequences or potentially driven by other factors.} In addition, given that the underlying phenotype model is known to be linear in this simulation study, these two kernels are the optimal choices by design. One should observe that using a linear kernel for GP regression corresponds exactly to Bayesian linear regression. 

For the regression task, we use an exact GP regression, while for binary traits we use a variational GP with probit likelihood \citep{opper2009variational}. The GP models are trained on a training set containing 80\% of the samples; the remaining 20\% is the test set. \new{Prior to training the GP models we centre to zero mean and scale to variance 1 the abundance of each OTU using the training samples only. The training mean and variance are used to centre/scale the test abundances prior to scoring.} Note that the three variants of the proposed string-based kernel are considered together with hyperparameters selected by maximising the training objective: the log-marginal likelihood (LML) for GP regression and the evidence lower bound (ELBO) for the variational GP. The optimised objective is used to evaluate the model fit alongside the log-predictive density (LPD) on the test set. 

\begin{figure}
	\centering
	\begin{subfigure}[b]{\columnwidth}
	    \includegraphics[width=\columnwidth]{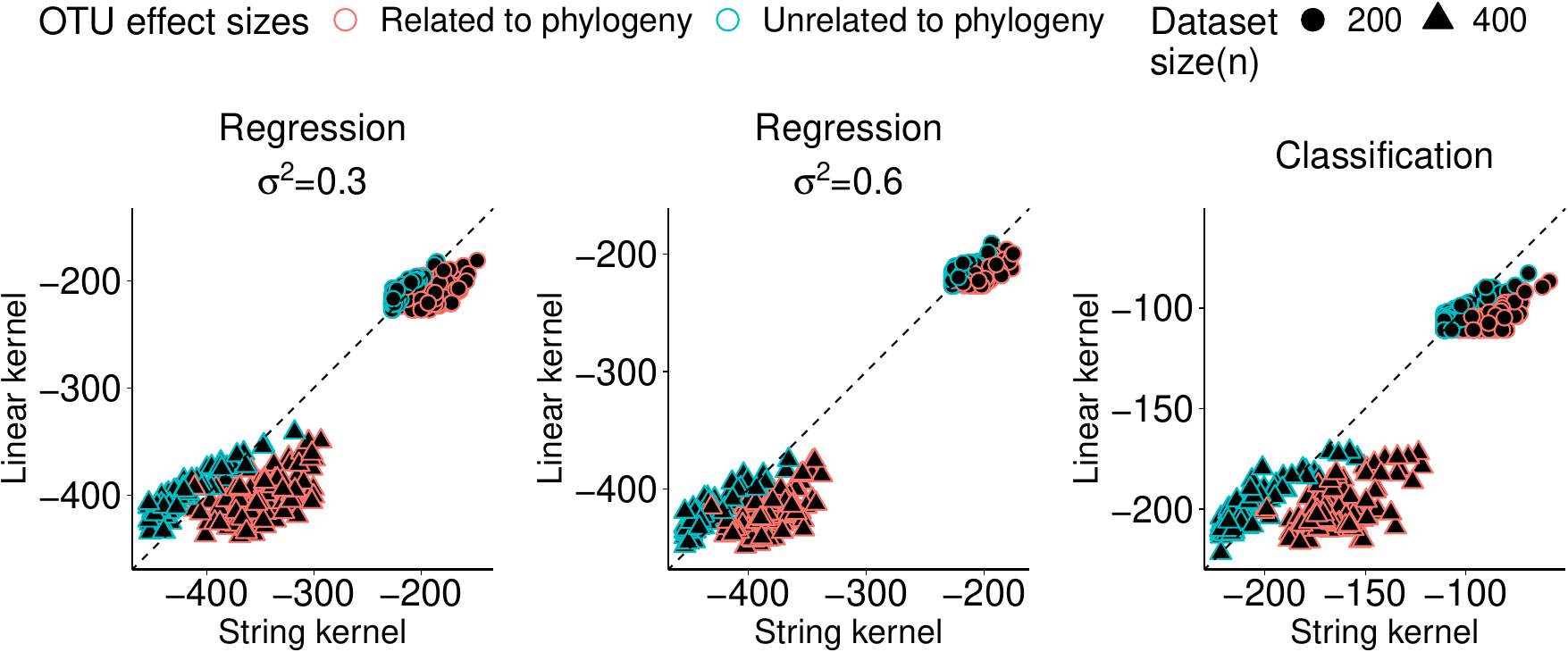}
	    \caption{Training objective (LML or ELBO)}
	\end{subfigure}
	\par\bigskip
	\begin{subfigure}[b]{\columnwidth}
	    \includegraphics[width=\columnwidth]{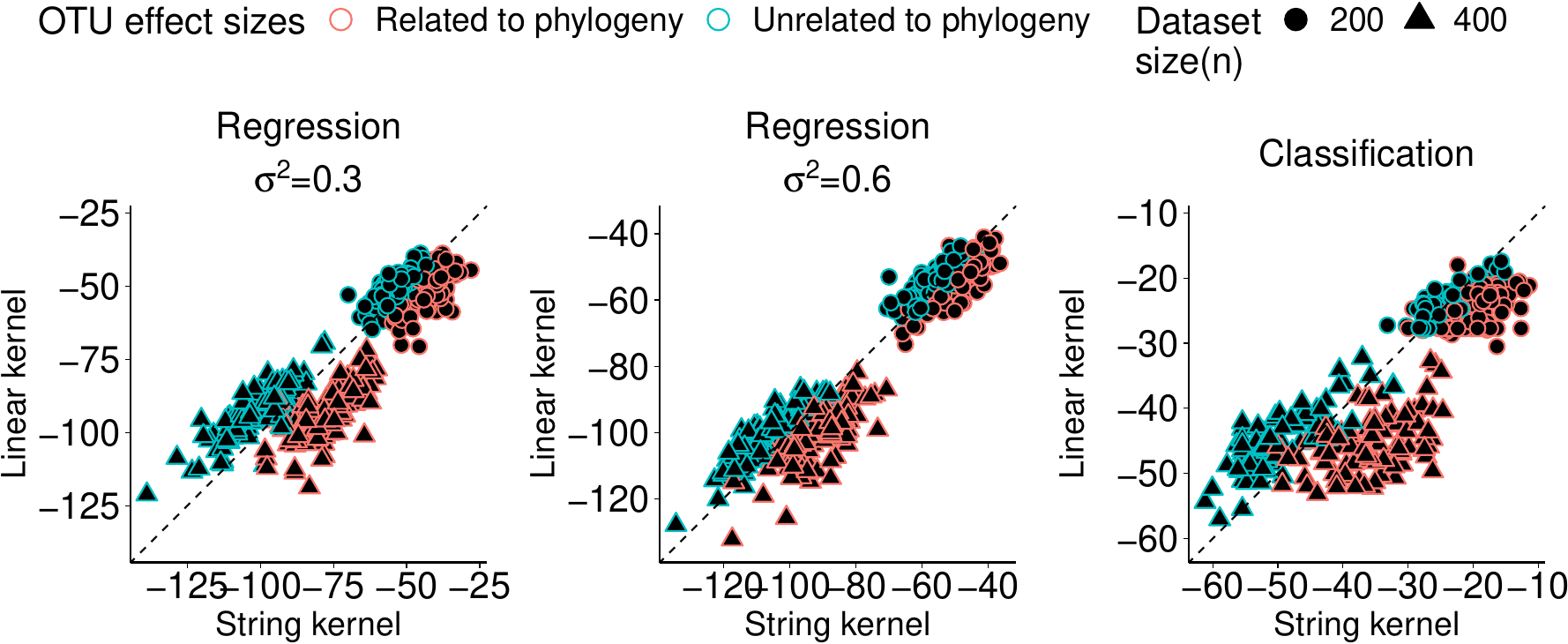}
	    \caption{Log-predictive density (held-out)}
	\end{subfigure}
	\caption{(A): training objective (LML for GP regression models and ELBO for the variational GP) for GPs with String and Linear kernels. Red dots correspond to datasets simulated under Scenario 1 where OTUs effect size are driven by the 16S rRNA gene sequence while blue dots correspond to datasets where effect sizes are unrelated to the phylogenetic tree. (B): The corresponding log-predictive densities show similar behaviour.}
	\label{fig:lmls_string_linear}
\end{figure}

Figure \ref{fig:lmls_string_linear} shows the difference in training objective (LML or ELBO) between GP models with a linear kernel and for the proposed string-based  kernel for each of the simulation setups. We observe that when the OTU effect size are not related to the phylogeny, both the string-based kernel and the linear kernel provide similar fit. However, when the effect sizes are related to the phylogeny, as expected, the string kernel provides a better fit in terms of both training objective and LPD as it does incorporate information on evolutionary similarities between OTUs. We therefore suggest to use the difference between the training objectives of a GP with a linear model and a GP with a string kernel 
to identify whether the factors controlling a host trait are related to the observed 16S rRNA gene sequence or if they are driven by other factors (such as areas of the bacterial genome that have not been sequenced or host/environmental factors). 
Note that, in the regression case the difference between the LML of two models is a Bayes factor, while for classification the ELBO can be used analogously for model selection \cite{cherief2019consistency}. 


\subsubsection{Impact of the string kernel hyperparameters in Gaussian process models}

\begin{figure}
    \centering
    \begin{subfigure}{\columnwidth}
        \centering
        \includegraphics[width=\columnwidth]{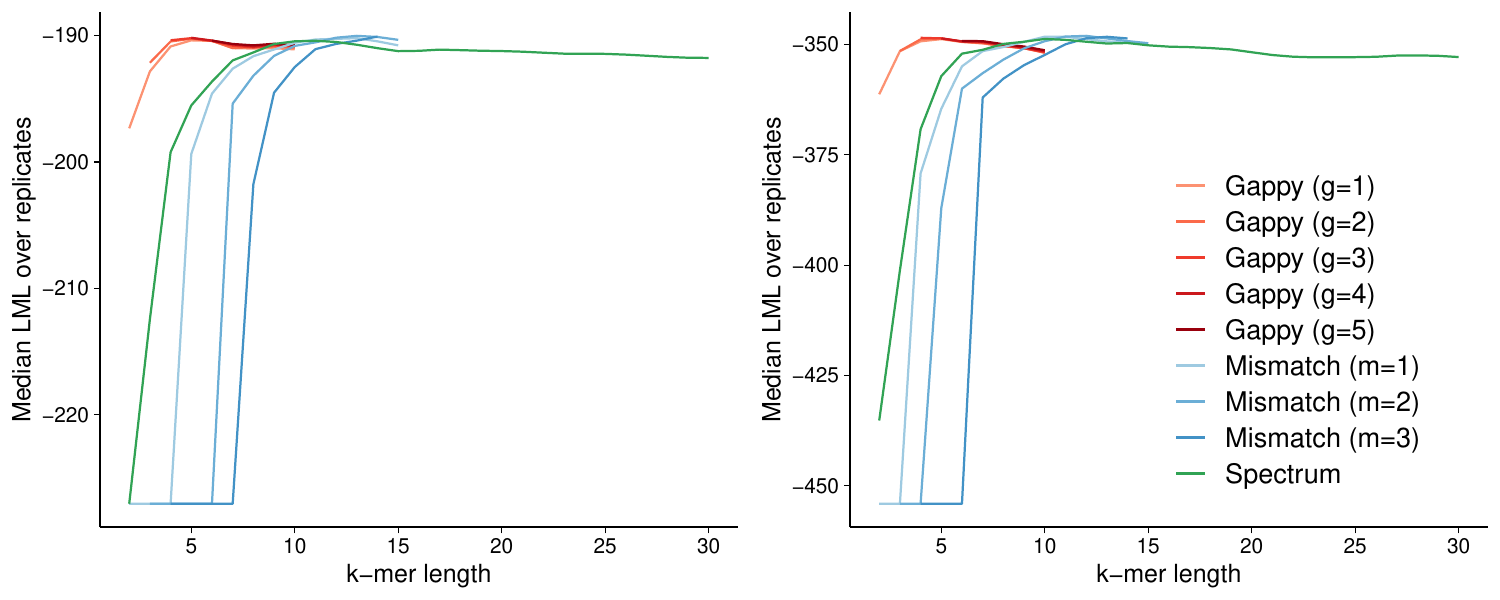}
        \caption{Regression with $\sigma^2=0.3$, $n=200$ (left) and $n=400$ (right)}
    \end{subfigure}
    \begin{subfigure}{\columnwidth}
        \centering
        \includegraphics[width=\columnwidth]{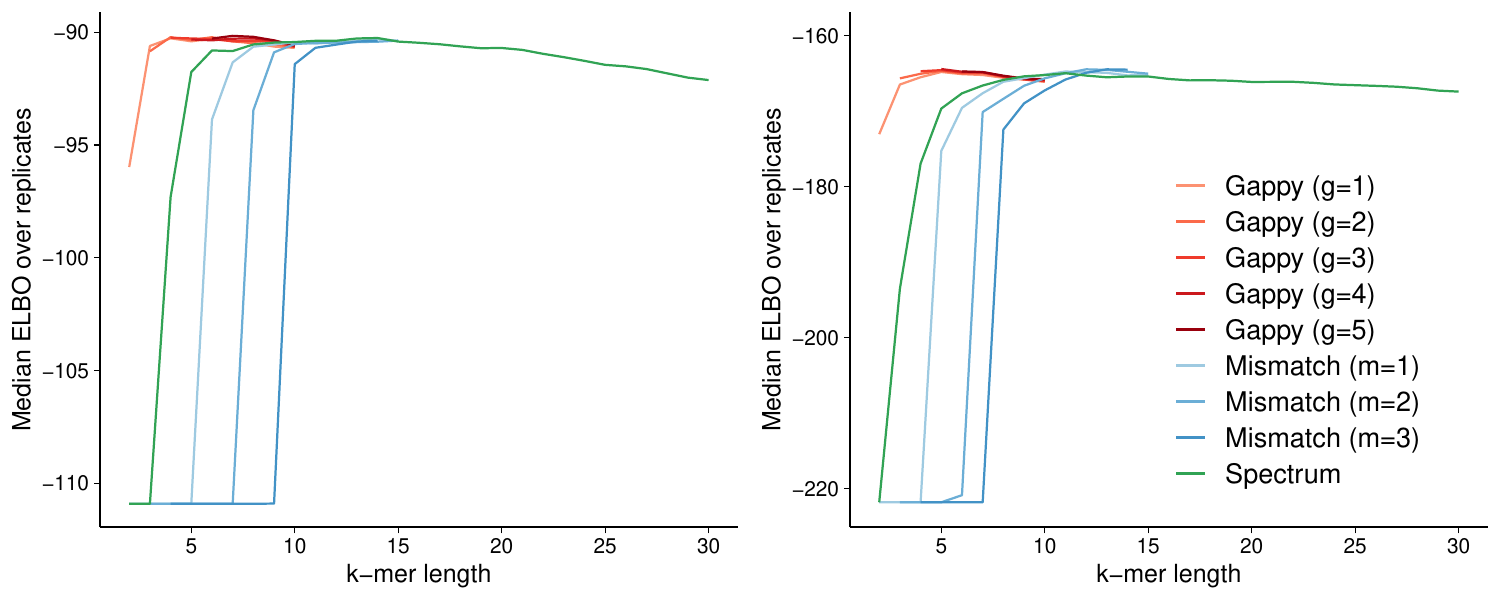}
        \caption{Classification, $n=200$ (left) and $n=400$ (right)}
    \end{subfigure}
    \caption{(A): Median log-marginal likelihood over 100 GP regression simulations. (B): Median ELBO over 100 GP classification simulations.}
    \label{fig:lml_and_elbo_string_hparam_dependence}
\end{figure}

We now present an investigation of the effect of the string-based kernel variant (Spectrum, Gappy pair or Mismatch) and the corresponding hyperparameters in Scenario 2 of the GP simulations (where OTU effects are driven by the 16S rRNA gene sequence). Figure \ref{fig:lml_and_elbo_string_hparam_dependence} shows the median log-marginal likelihood for the regression case and the ELBO for the classification for GP models with different string kernel hyerparameters across the 100 simulation replicates. A general trend is observed: low values of $k$ results in poor performance, while increasing the $k$-mer length initially leads to significant improvements. Performance then peaks at an optimal $k$ before plateauing and gradually declining as $k$ continues to increase. For the Gappy pair kernel, there is a weak dependence on the number of gaps, $g$, while for the Mismatch kernel, the value of $m$ has a significant impact only when the length of the $k$-mer is small. Notably, for any given $k$, the Gappy pair kernel consistently achieves a higher LML or ELBO compared to the other kernel variants.



\subsection{Real data applications - host trait prediction} \label{sec:gp_real_data}

We now demonstrate the use of the proposed string-based kernels on a host-trait prediction problem from a real dataset. This is a regression task with $n=388$, $p=525$ predicting vaginal pH from bacterial community composition \cite{ravel2011vaginal}. In this dataset the sequences are clustered to 100\% identity and so are termed amplicon sequence variants (ASVs) rather than OTUs, which are clustered to 97\% identity.

\begin{figure}[h]
    \captionsetup[subfigure]{justification=centering}
	\centering
	\begin{subfigure}[t]{0.45\columnwidth}
		\centering
		\includegraphics[width=\textwidth]{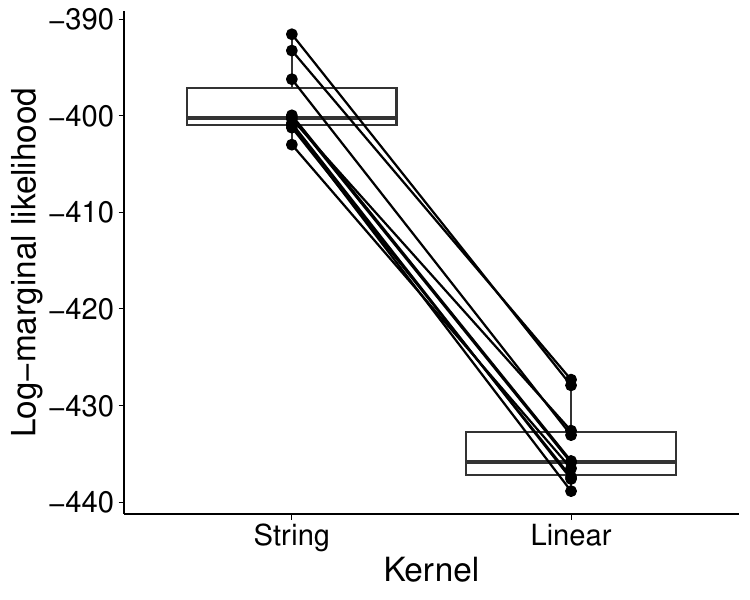}
		\caption{Log-marginal likelihood on training folds}
	\end{subfigure}
	\hspace{0.1cm}
	\begin{subfigure}[t]{0.45\columnwidth}
		\centering
		\includegraphics[width=\textwidth]{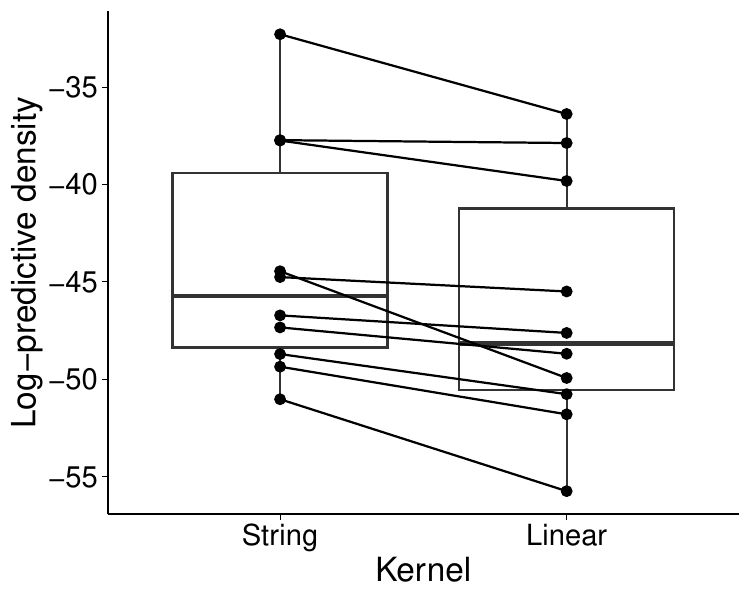}
		\caption{Log-predictive density on held-out folds}
	\end{subfigure}
	\hspace{0.1cm}
	\caption{Real data application of host trait prediction using a GP: predicting vaginal pH from vaginal bacterial community composition \cite{ravel2011vaginal}. Log-densities are estimated using ten-fold cross-validation.}
	\label{fig:real_data_results}
\end{figure}

We used ten-fold cross-validation to estimate the log-marginal likelihood and log-predictive density on the training and held-out samples respectively. \new{For string kernel model selection we optimised the log-marginal likelihood separately for each hyperparameter combination and selected the model with the largest optimised LML value.} In each iteration of cross-validation, we trained two GP models, one with a String kernel and one with a Linear kernel, and compared the resulting two log-marginal likelihoods (or log-predictive densities). As discussed in the previous Section, this analysis can be framed as a comparison of two competing biological hypotheses: whether the associations between taxa abundance and vaginal pH are correlated or uncorrelated with phylogenetic similarity. The log-marginal likelihoods from each fold are shown in Figure \ref{fig:real_data_results}(A), which indicates that the GP model with a String-based kernel clearly provides a better fit than when using a Linear kernel. This also corresponds to better predictive performance (higher log-predictive density) on the held-out data (see Figure \ref{fig:real_data_results}(C)). These results therefore support a correlation between taxa effect on vaginal pH and phylogenetic similarity in this dataset. 


\section{Discussion and Conclusion} \label{sec:discussion}



Our work introduces a novel family of kernels for microbiome datasets that leverages the fact that each Operational Taxonomic Unit (OTU) is defined by a representative DNA sequence. Unlike traditional approaches that rely solely on taxonomic abundance, our proposed kernel family encodes the similarity between OTUs by applying string kernels—commonly used in natural language processing—to their representative sequences. This framework constructs an inner product space based on string kernel similarities, which is then used to compute sample-wise similarity. Our results demonstrate the utility of this approach in incorporating phylogenetic information into two key tasks: (i) the kernel two-sample test and (ii) host-trait prediction using Gaussian Processes (GPs). In the remainder of this section, we summarise our key findings, discuss the limitations of our approach, and outline directions for future research.\\

Our two-sample test simulations revealed that commonly used characteristic kernels, such as Linear or RBF kernels, may be inadequate for analysing 16S rRNA gene sequencing data, at least under the assumptions of our study. We considered scenarios where differences between the distributions $P$ and $Q$ occurred through permutations of the underlying $\alpha$, when there are many other ways for two populations to differ. However, this simulation setup was constructed to demonstrate the undesirable behaviour of the abundance-only kernels in this setting, as well as show that the proposed kernels do not exhibit these behaviours. This aim was achieved and these findings are sufficient to warn against using kernels that ignore phylogenetic relationships between samples in a two-sample test on OTU-level data (or at least to exercise caution when performing such tests).

Moreover, our simulation results showed that a kernel two-sample test using one of the proposed string-based kernel demonstrates the desirable property of being sensitive to the phylogenetic scale (denoted by $\varepsilon$) at which the difference between the two probability distributions $P$ and $Q$ occur. However, a systematic method for tuning the string kernel hyperparameters to be sensitive to a desired value of $\varepsilon$ remains an open problem and would be an interesting avenue for future research.

This study focused on the kernel two-sample test as proposed by Gretton et al \cite{gretton2012kernel}, which uses MMD as the test statistic, and host trait prediction using GP models. \new{A natural extension of our approach is therefore to investigate the performance of the proposed string-based kernels in other kernel-based microbiome analysis methods wherever RBF kernels are the default choice. This includes prediction using kernelised support vector machines \cite{li2025best,topccuouglu2020framework,nguyen2021associations,ghannam2021machine}. Other approaches (including MiRKAT and its extensions), already utilise the UniFrac kernel to model phylogenetic relationships but an investigation of how these methods perform with our proposed kernel will still be of interest.}


The host trait prediction simulation study showed that the GP training objective -- either log-marginal likelihood (LML) or ELBO -- of GP models using one of the proposed kernel vs a linear kernel can be used as an indicator of the distribution of OTU effects on host phenotype across the phylogenetic tree. As the tree is constructed from the 16S rRNA gene sequences this summary statistic therefore quantifies the degree to which the OTU effects are explained by 16S rRNA gene sequence variation. If a GP with a linear kernel has a larger LML than one with a string kernel then the OTU effects must be explained by (i) variation in parts of the microbial sequence that have not been collected or (ii) by non-sequence (e.g. environmental) factors.

However, this approach has only been shown to be effective when the assumptions of the simulation are met. The most important of these is that the host phenotypes depends linearly on the relative abundance. An interesting option for future work is to investigate the robustness of the results to mis-specification of the phenotype model (when the phenotype model contains non-linear dependencies but the phylogenetic kernel remains linear). However, one of the benefits of GPs is their modularity and so it is straightforward to combine string and characteristic kernels to model both phylogeny and nonlinear effects. One way to achieve this is to use the following kernel: $$k_q(x,x') = \exp \left( -(x-x')^T S_q (x-x') \right)$$ which model non-linear dependencies between abundances while incorporating phylogenetic similarities between OTUs via the matrix $S_q$.\\

A final limitation of these experiments is that they focus on modelling the phylogenetic relationships amongst the OTUs and have largely neglected some other important features of OTU count data: sparsity and zero-inflation. While the simulation setup ensured these features were present in the simulated OTU tables they were not explicitly modelled by the kernel two-sample test nor the GP models. The aforementioned modularity of GPs also enables the construction of a GP that models both zero-inflation of counts and phylogenetic relationships combining kernels. This modularity is one of the reasons why kernel methods are a popular approach for biological data integration as their additive and multiplicative properties enables the straightforward combination of heterogeneous data types \cite{daemen2009kernel,heriche2014integration,mariette2018unsupervised}.

\new{A final limitation of our approach is that it relies on the assumption that the similarity between representative sequences is a good proxy of evolutionary distance between OTUs. This is a limitation shared by any phylogenetic analysis of such datasets, as alternative methods (e.g. UniFrac) also utilise these sequences to construct phylogenetic trees. While this approach has allowed for cost-effective identification and quantification of bacterial communities, there are known limitations to using $\sim$200 base pair sub-regions of the 16S rRNA gene, such as limited taxonomic resolution at species level \cite{johnson2019evaluation}. Better taxonomic resolution can be achieved by sequencing the entire 16S rRNA region \cite{jeong2021effect} or by shotgun sequencing of the entire bacterial genome \cite{durazzi2021comparison}, but the majority of studies still target a carefully selected subset of the 16S rRNA gene \cite{lopez2023determining}. Given that our approach only requires representative sequences it can be applied without modification to newer datasets utilising these improved technologies.}






\section{Abbreviations}

\begin{itemize}
    \item ELBO: evidence lower bound
    \item GP: Gaussian process
    \item LPD: log-predictive density
    \item rRNA: Ribosomal ribonucleic acid 
    \item LML: log-marginal likelihood
    \item OTU: operational taxonomic unit
    \item MMD: maximum mean discrepancy
    \item RBF: radial basis function
    \item DMN: Dirichlet mulitnomial
    \item CLR: centre log-ratio
    \item RKHS: Reproducing kernel Hilbert space
\end{itemize}


\section*{Acknowledgements}

Jonathan Ish-Horowicz was the recipient of a Wellcome Trust PhD studentship (215359/Z/19/Z).

 \bibliographystyle{elsarticle-num} 
 \bibliography{cas-refs}





\end{document}